\RequirePackage{fix-cm}

\documentclass[smallextended]{svjour3}
\usepackage{hyperref}
\usepackage{lscape}
\smartqed  
\usepackage{multirow}
\usepackage{natbib}
\usepackage{graphicx}

\journalname{Quantum Information Processing}

\begin{document}

\title{Heterogeneous Quantum Computing for Satellite Constellation Optimization: Solving the Weighted K-Clique Problem}
\author{Gideon Bass \and Casey Tomlin \and Vaibhaw Kumar \and Pete Rihaczek \and Joseph Dulny III}
\institute{Booz Allen Hamilton\\
		901 15th Street\\
		Washington DC, 20005 USA}

\date{Received: July 6 2017 / Accepted: July 6 2017}

\maketitle

\begin{abstract}
NP-hard optimization problems scale very rapidly with problem size, becoming unsolvable with brute force methods, even with supercomputing resources. Typically, such problems have been approximated with heuristics. However, these methods still take a long time and are not guaranteed to find an optimal solution. Quantum computing offers the possibility of producing significant speed-up and improved solution quality. Current quantum annealing (QA) devices are designed to solve difficult optimization problems, but they are limited by hardware size and qubit connectivity restrictions. We present a novel heterogeneous computing stack that combines QA and classical machine learning, allowing the use of QA on problems larger than the hardware limits of the quantum device. These results represent experiments on a real-world problem represented by the weighted $k$-clique problem. Through this experiment, we provide insight into the state of quantum machine learning.

\end{abstract}	

\section{Introduction}
Optimization problems, where a particular objective function must be extremized, have widespread applications in a variety of fields. In this paper, we apply quantum annealing to solve one particular example: Grouping satellites in order to maximize coverage. This problem is interesting both for its direct application, and because, as we transform it to the more general weighted $K$-clique problem, the techniques described are applicable to many other optimization problems in other domains.

This problem and many others are NP-hard, and scale very rapidly with problem size. In such cases, they quickly become unsolvable with brute force methods, even with supercomputing resources. We may not be able to rely on Moore's Law as it slows while approaching physical limits \citep{Kumar_2015}, and problems which scale exponentially or worse will remain intractable even given centuries of Moore's Law-like growth \citep{Garey_1990}. Therefore, hard optimization problems have typically been treated with heuristics. Heuristics can still take a long time to reach satisfactory answers, and are not guaranteed to produce the optimal solution. Quantum annealing \citep{Brooke_1999,Farhi_2001,Finnila_1994,Kadowaki_1998,Santoro_2002,Santoro_2006} is a computing technique suited to solve NP-hard combinatorial optimization problems \citep{Farhi_2001,Garey_1990}. In quantum annealing, a system of qubits is initialized in the easy-to-prepare €ground state of a `driver' Hamiltonian. Adiabatic evolution of the driver Hamiltonian into a `€œproblem'€ Hamiltonian (encoding the optimization problem to be solved) sends the initial state to the ground state of the problem Hamiltonian, such that the state after adiabatic evolution encodes the solution of the optimization problem. In particular, the adiabatic theorem \citep{Born_1928, Kato_1950} gives conditions on the evolution such that the system remains in the instantaneous ground state with high probability.

Current quantum annealers solve quadratic unconstrained binary optimization (QUBO) problems, also known as Ising spin-glasses. However, only a limited number of couplings between binary variables can be set due to hardware limitations.

While maturing rapidly, QA is still ``€œbleeding-edge'' technology, and thus is not ready to supplant traditional computing. In particular, the hardware is currently limited in size (around 1000 qubits), connectivity (maximum of six connections per qubit), and precision (currently around 5 bits) \citep{D-Wave2015}. With these limitations, it is difficult to fully embed real-world optimization problems on the device.  If these limitations could be overcome, there is strong evidence from experiments \citep{Denchev_2015} and simulations \citep{Crosson_2016} that QA should produce substantial speedups. QA has been applied to real problems in diverse areas such as multiple query optimization \citep{Trummer_2016} and prime factorization \citep{Dridi_2016}.

We therefore explore the use of quantum annealing in a heterogeneous, classical-quantum approach, taking advantage both systems' strengths. This is a ``€œbest of both worlds,'' synergistic approach where we use the large size of classical computers to analyze the problem as a whole, supplemented by QA to evolve directly into the ground state. We explore the ability of this heterogeneous approach to showcase its potential and produce significant performance improvements.

\subsection{Problem Statement}
In order to test the viability of this schema, we have worked on the real-world optimization problem of splitting a set of satellites, called a constellation, into further small groups, or sub-constellations. In this problem, a single constellation of $N$ satellites are to be divided into $k$ sub-constellations, with $N$ and $k$ pre-determined based on availability. The goal is then to find the assignment of each satellite to a sub-constellation such that the total coverage of a designated Earth region is maximized. Coverage for arbitrary sub-constellations can be efficiently calculated as the percentage of time that the Earth region of interest is within the signal range of at least one satellite, and this data is then provided to our algorithm in the form of a lookup table.

This problem can be expressed as a weighted $k$-clique optimization. In graph theory, a graph consists of a set of vertices and edges connecting pairs of vertices. A clique is a subset of a graph where every two distinct vertices in the clique are connected by an edge, with a $k$-clique consisting of $k$ such vertices.  Each grouping of satellites can be represented as a vertex in a graph, with the weight of the vertex equal to the coverage provided by that grouping. Groupings that do not both use the same satellite (which would be an invalid assignment) are vertices which are connected by an edge. A clique then consists of a set of mutually exclusive satellite groupings, i.e. a unique partitioning of the full constellation into independent sub-constellations. The goal then is to find the set of $k$ connected vertices (independent sub-constellations) that maximizes the weights of those vertices. 

\begin{figure}
	\centering
	\includegraphics[width=0.7\linewidth]{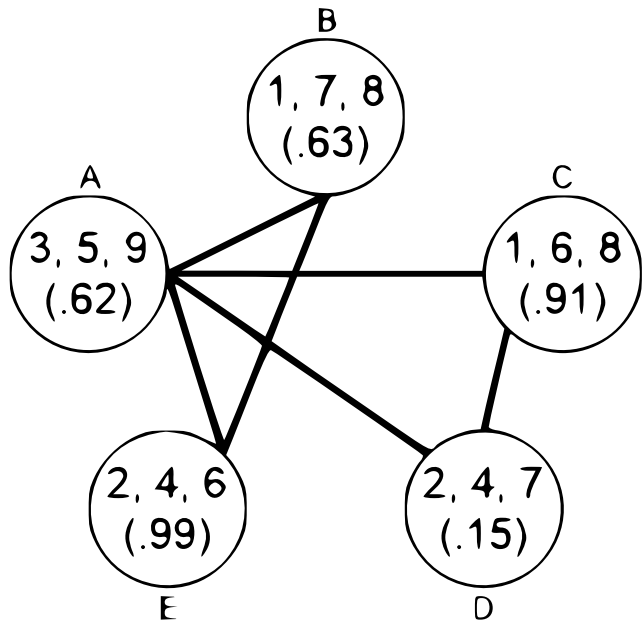}
	\caption{Example of $k$-clique optimization graph with 9 satellites being divided into three groups.}
	\label{fig:ConstellationExample}
\end{figure}

As a simplified example, see figure \ref{fig:ConstellationExample}. Here, we have a simplified 9-satellite problem. Each vertex represents one of the many possible groups of 3 satellites (which three are denoted by the first three numbers), as well as the resulting coverage (indicated by the fourth number in parenthesis). The lines represent edges between vertices, i.e. vertices which do not use the same satellite. For example, there is no connection between vertices B and D, because they both use satellite 7. 

In this simplified case, there are two cliques: ABE, and ACD. ABE has a total average coverage of 0.75 (average of 0.62, 0.63, and 0.99), while ACD has an average coverage of only 0.56 (average of 0.62, 0.91, and 0.15). Thus, if these were the only options, ABE would be the ideal case. Note that in the real-world problem, there would be additional vertices representing other combinations; these have been omitted to simplify the diagram. 

\subsection{Classical Approach}
The coverage data can be viewed as an implicit graph, meaning that the edges can be calculated from the vertex data itself. Each vertex represents a unique combination of satellites. For computational purposes, this is represented by a binary bitstring. For example the binary representation 1001101011010001 read from right to left would represent satellite numbers 1, 5, 7, 8, 10, 12, 13, and 16 being in the sub-constellation, with each such unique grouping having an associated coverage percentage as its weight. This is helpful because if the bitwise AND of two sets of satellites, a very quick computation to check, is zero, the sets have no satellites in common and thus the vertices share an edge. A group of k mutually independent sub-constellations is then a k-clique in the graph. However, with typical problem sizes ($25<N<35$ and $5<k<10$), the size and density of the resulting graphs poses significant challenges to classical computing. 

Assume the full constellation consists of 40 satellites. A single vertex representing an 8-satellite sub-constellation would then leave 32 satellites not in the set, which could then be divided into ${32 \choose 8} = 10,518,300$ other 8-satellite vertices with which it shares an edge, ${32 \choose 7} = 3,365,856$ 7-satellite vertices with which it shares an edge, etc. In other words there can be millions of vertices, each having millions of edges with other vertices. The general problem calls for partitioning n satellites into k groups, and the number of ways this can be done, i.e. the number of k-cliques, is given by the Stirling set number formula $S(n, k)$. A constellation of 40 satellites can be split into 5 groups $S(40, 5)$ = 7.57409e+25 ways, meaning this many 5-cliques need to be found and evaluated to prove the global optimum. Therefore certain heuristics must be employed to reduce the search space. 

First, it can be observed that coverage is correlated with the number of satellites in a sub-constellation, and so groups of only one or two satellites can be ignored. More simply, an arbitrary minimum coverage number such as 85\% can be used to eliminate low-weight vertices and cliques. It is further empirically observed that solutions where the number of satellites is evenly distributed into sub-constellations tend to have the best results. This may be because increasing the size of some sub-constellations at the expense of others results in diminishing returns as there is increasing overlap in coverage in the larger sets, thereby wasting resources. 

Because of the density of the graph, there is a high probability of finding higher-weight cliques in the neighborhoods (the set of vertices connected to a given vertex) of high-weight vertices. So the vertices are first sorted in descending coverage order, and then a parallel k-clique search is performed which prioritizes high-value vertices and their neighborhoods. The sorting also facilitates various pruning techniques, where branches of the search can be ignored when the remaining vertices in a neighborhood could not be added to an existing clique without exceeding the total number of satellites in the full constellation, or when there is no mathematical possibility of finding a higher-weight clique than the current best already found by any thread.

Using such heuristics, which often depend on some knowledge of the underlying physical problem, it is possible to find acceptable solutions in minutes or even seconds depending on the exact problem definition and coverage inputs, but proving a global maximum could still take hours, days, or much longer. An additional worry is that the problem space scales rapidly with the number of satellites and groups. With even minor increases in numbers, this heuristic pruning-based approach may fail. Therefore, there is keen interest in determining whether a quantum annealing-based approach may offer performance improvement in less time, and determining how different problem sizes scale. 

\section{Methods}
In order to use current hardware quantum annealing devices to treat our problem, it must be recast as a QUBO, which we detail in the following subsections.

\subsection{Overview of the QUBO formulation}
The goal in quadratic unconstrained binary optimization is to (in this case) minimize the `energy' expression (the problem Hamiltonian in the context of quantum annealing)

\begin{equation}
H_{\mathrm{P}} = \sum_{i} h^i q_i + \sum_{(i,j)} J^{ij} q_i q_j, \label{eq:quboham}
\end{equation}
as a function of the set of binary variables $\{q_i\}$, given the set of biases $h^i$ and couplings $J^{ij}$ as external parameters.

In a true QUBO problem, there is no restriction on the values of $h$ and $J$, but when employing physical hardware with limited connectivity \citep{Bunyk_2014}, many $J$ values are forced to vanish, and the precision of both $h$ and $J$ are limited. Connectivity restrictions can be avoided through use of redundant chained qubits that yield more highly-connected logical qubits at the price of fewer available computational qubits. We use the D-Wave heuristic embedding solver \citep{embeddingz} to pre-search for a fully-connected embedding for the chip, and then use that embedding for all quantum annealing computations. 

\subsection{QUBO for coverage optimization}
This satellite coverage optimization problem can be described in terms of graphs, where each vertex is a potential set of satellites, and connections (edges) are created when sets do not use the same satellite. In this formulation, translating into QUBO form is relatively simple. There are essentially two constraints:

\begin{enumerate}
\item Choose a specific number of vertices such that all are connected.
\item While obeying constraint 1, choose the vertices that have the highest total sum of the individual coverages.
\end{enumerate}
For the first constraint, we first reverse the problem, preferentially rejecting any pair of vertices that use the same satellite. In the terms of the QUBO, each coupling $J$ between two overlapping vertices is assigned a very large penalty. This prevents the solution from using the same satellite twice. This penalty term is

\begin{equation}
H = \sum_{i<j} 2(w_i + w_j) (x_ix_j)
\end{equation}
if $i$ and $j$ use the same satellite, and zero otherwise, and where $w_i$ is the weight on node $i$ that represents the coverage of that vertex. This penalty is set at this value because of the following reasoning: In the worst case, imagine that there are two sets of satellites that overlap, $i$ and $j$. By turning each one on, per equation \ref{eq:quboham} we would decrease the energy by an amount equal to $w_i+w_j$. Thus, by adding twice that amount, we ensure that there is a net increase in energy and this choice is preferentially rejected. Of course, if group $i$ or $j$ overlaps with other activated groups.

The second part of the first constraint, that a specific number of vertices be chosen, is trickier. If there are $k$ sub-constellations, there will be $k(k-1)$ penalty terms, and these terms encourage the use of low-weight sub-constellations. E.g. for $k=8$ (a configuration of eight groups), there are 56 penalty terms, each of weight $\approx w/8$ where $w$ is the average fitness of the groups. On the other hand, a configuration of seven sub-constellations will have 42 penalty terms, each of weight $w'/8$ where $w'$ would (in principle) be a bit larger than w from distributing the extra vehicles across the sub-constellations. So the reduction in penalty would be 7w - 5.25w$\prime$ which is still greater than w, while the loss in fitness by removing the sub-constellation with the least coverage is no worse than w, so it looks energetically favorable to have fewer sub-constellations.

This kind of penalty model is not guaranteed to produce configurations of k sub-constellations, but with some tweaking, was generally found to produce results that worked most of the time. However, some classical post-processing was required to fix up mistakes.

In the end, to produce \textbf{k} sub-constellations we used the penalty model: 
\begin{equation}
\label{eqn:PenaltyModel}
H = W \Bigl( \sum_i x_i - k \Bigr)^2 =k^2W - \sum_i 2kW x_i + \sum_{i<j} x_i x_j
\end{equation}
where $W$ is the max-weight, and the first term is a constant that is absorbed into the energy offset and can be safely ignored. 

For constraint 2, each h term is assigned a value directly proportional to the coverage amount for that particular grouping. This produces the simple Hamiltonian:
\begin{equation}
H=\sum_i -A w_i x_i
\end{equation}
Where again $w_i$ is the coverage of one particular node. In theory, the linear constant $A$ should be set equal to 1. However, because $w_i$ is at most equal to $W$, so long as any linear increase to A is balanced by a 1 smaller decrease to the linear terms from the sub-constellation group above (in equation \ref{eqn:PenaltyModel}) the constraints are not violated. Essentially, increasing A increases the importance of the minimization constraint, relative to decreasing the importance of the number of sub-constellation constraint.
The final, combined Hamiltonian is then:
\begin{equation}
H=\sum_i \Bigg(-Aw_i-\bigg(2k-(A-1)\bigg)W\Bigg)x_i + \sum_{i<j} \bigg(2W + 2O_{ij}(w_i+w_j)\bigg)(x_ix_j)
\end{equation}
Where $O_{ij}$ is 1 if $i$ and $j$ share at least one satellite, and zero otherwise.

The relative value of $A$ was first explored in early work on this problem, and we empirically found, at least for this instance, that a value of 4 produced the best results, and this value was used for all of the results described below. Above this value, the number of sub-constellation constraint began to be violated regularly, while lower values gave solutions with lesser coverage. Further work exploring possible values of $A$ while the other hyperparameters described below are also varied would be of great interest, as would an enhanced theoretical understanding of why the constraint violations begin to occur where they do.

\subsection{Classical Co-Processing}
\subsubsection{Genetic Algorithm Preprocessing}
Having produced a QUBO formulation, the next problem was the size of the problem. Given typical values of n=31, k=8, and allowable groups of size anywhere from 3 to 8, there are a little over 3.5 million possible groups. Given that currently available quantum annealers have 1000-2000 qubits, this QUBO cannot possibly be fully encoded into the annealing device, and likely will be beyond the capabilities for many years, even if a Moore's Law like growth continues for some time. Further complicating matters, the QUBO above is fully connected, while current devices have only limited connections. Multiple qubits can be "chained" together, forming logical qubits that have greater connectivity, but this comes at a reduced number of computational qubits. All told, this problem would seem to be impossible for the foreseeable future.

In order to begin attempting this problem, we used a classical heuristic solver, a genetic algorithm (GA), to explore the solution space. GA's use the principles of biological evolution to evolve from an initial "population" of poor solutions and, through selection of the "fittest" individuals, produce steady progression towards the true optimum. 

For the purpose of this problem, GA's have two very helpful properties. First, they explore the entire solution space in a fairly efficient but random way. By running on a classical computer, small selections of the million-plus possible groups can be chosen. Second, they can be run for an arbitrary amount of time, after which a population can be found that contains a large number of possible solutions. We can then use the groups (not sets of groups) that occurred most frequently in this final population to run in the quantum annealing device, choosing the largest possible amount that can be encoded and embedding into our quantum hardware. 

\subsubsection{Other Preprocessing Methods}
As a comparison, we also attempted two other more naive pre-processing techniques. These were random selection and pruned selection. In random selection, a random number of possible groups was chosen. To ensure there were a reasonable number of small groups (which are a smaller percentage of the total number of groups), the random choice was performed per group size, with equal amounts of groups of size 3, 4, etc. 

Next, as mentioned in the classical processing section, it is known that for this particular problem, pruning to only select the best individual groups produces excellent results. We thus performed this pruning, again ensuring that equal numbers of each size group was chosen. 

\subsubsection{Classical Post-Processing}
Quantum annealing is an inherently random process, and it is quite likely that the solution produced, even after many runs, will be close to, but not exactly the true global minimum. In some cases, the constraints will be broken, resulting in too many or too few groups chosen or in multiple groups using the same satellite being selected. 

To avoid this, after the quantum annealing run, we added a greedy classical post-processing stage. In this stage, if there are too many groups, we iteratively remove the one with the lowest coverage. If there are too few groups, we iteratively add groups of size 3 using satellites not currently in use. Then, any satellites that are being used in multiple groups are removed from whichever group results in the least decrease in coverage. Finally, if there are any unused satellites, they are each individually added to the group that would result in the greatest increase in coverage.

While this process is by no means guaranteed to produce a true global minimum, it is guaranteed to produce a legal solution without decreasing the coverage from the QA solution. 

\subsection{D-Wave hyperparameters}
The D-Wave device itself, while it performs only one type of computation, does have several hyperparameters that allow limited control in how the annealing process works. The primary ones that we investigated are annealing time, number of reads, programming thermalization, reading thermalization time, and number of spin reversals. There are also two D-Wave solvers and an optional post-processing step.

\subsubsection{AnnealTime}
There is a default time for the annealing cycle. This parameter allows the user to change that default value. For example, on the DW2X, the lower limit is 5 $\mu$s, and the upper limit is 2000 $\mu$s.

\subsubsection{NumReads}
When a specific problem is loaded, the QPU can run multiple independent "annealing cycles." For each complete cycle, a separate answer is returned, enabling detection of statistical patterns and increasing the chance of finding the true global minimum. 

\subsubsection{ProgTime}
As part of the annealing process, the processor must wait for a short time after programming the processor, in order to cool down to the base temperature. Lower values will speed up solving at the expense of solution quality. On the DW2X QPU, the lower limit is 1 $\mu$s, and the upper limit is 10,000 $\mu$s.

\subsubsection{ReadTime}
Similar to programming time, after the annealing process there is a mandatory wait time before the solution states can be read from the processor, in order to cool
down to the base temperature. 

\subsubsection{Solver}
New in the DW2X, D-Wave introduced the "Virtual Full Yield Chip" (VFYC). All DW2X chips have a few qubits and connectors that do not work, due to errors in the manufacturing process. This means that the same problem will be embedded differently into one D-Wave chip rather than another. To fix this problem, the VFYC mode allows a problem to be submitted formatted as if all qubits and connectors are functional. Classical heuristic post-processing supplements the D-Wave to determine the solution to this problem.

\subsubsection{PostProc}
Also new to the DW2X, there are multiple post-processing options. In the VFYC mode, the optimize post-processing is mandatory, but in regular DW2X, the user has the option of performing a post-processing optimization or sampling. The latter is used for deep-learning, and is not appropriate for this problem, so was not used. However, we did experiment on the effectiveness of the optimize post-processing.

\subsection{Exploring the hyperparameter space}
There were a total of 13 hyperparameters. Table \ref{tab:hyperparameters} gives a brief description of each hyperparameter, its range of values, and whether the hyperparameter is from our pre-processing methods or a D-Wave provided choice. 

To explore this hyperparameter space more fully, we performed approximately 1500 independent runs, with each hyperparameter chosen as a random value within the acceptable range. Note that some hyperparameters choices did depend on others, notably the GA hyperparameters NumGen, PopSize, and MutRate were only chosen when method was Genetic, and the D-Wave hyperparameter SpinRev is limited to less than the total number of repetitions. By choosing randomly, we are able to somewhat map out the full hyperparameter space with less computational resources than an exhaustive grid search would need.

We note that the choice of hyperparameters is itself an optimization problem. Because the solution space is a-priori completely unknown, this is not a good target for quantum annealing, but other optimization methods such as the genetic algorithm or simulated annealing could have been used in place of our random choice. For some classes of problem, where there may be a large time for training but afterwards the problem will need to be solved accurately and quickly, this approach might be of value. 

\begin{table}
\caption{All Hyperparameters} \label{tab:hyperparameters}
\begin{center}
\begin{tabular}{|p{15mm}|p{11mm}|p{12mm}|p{32mm}|}
	\hline 
	Name & Range & Type & Description \\ 
	\hline 
	Method& Genetic, Random, Prune & Pre-processing & Pre-processing method \\ 
	\hline
	NumGen& 10-1000 & Pre-processing & Number of generations for GA \\ 
	\hline  
	PopSize& 10-1000 & Pre-processing & Population size for GA \\ 
	\hline  
	MutRate& 0.01-0.25 & Pre-processing & Mutation rate for GA \\ 
	\hline  	
	NumNodes& 30-49 & Pre-processing & Number of nodes selected in pre-processing stage \\ 
	\hline  
	LargestGroup& 4-7 & Pre-processing & Largest set size used\\
	\hline
	NumReps& 10-10000 & D-Wave & Num of annealing repetitions\\
	\hline	 
	AnnealTime& 5-2000 & D-Wave & Annealing Time ($\mu$s)\\
	\hline	 
	ProgTime& 1-10000 & D-Wave & Input Thermalization time ($\mu$s)\\
	\hline	 
	ReadTime& 1-10000 & D-Wave & Output Thermalization time ($\mu$s)\\
	\hline	 
	SpinRev& 1-NumReps & D-Wave & Number of Spin Reversals\\
	\hline	 
	Solver& DW2X, VFYC & D-Wave & Solver Used, actual chip or virtual full yield\\
	\hline
	PostProc& Optimize, None & D-Wave & D-Wave Post-Processing\\
	\hline	 	
\end{tabular} 
\end{center}
\end{table}

\section{Results}
Experiments were performed on both QA simulators and early-stage commercial QA hardware. Early exploration was run using Quantum Monte Carlo (QMC) algorithm \citep{Isakov_2015} which cheaply simulates QA. This enabled us to easily explore QUBO formulations of the problem. 

Later, we performed D-Wave runs on a DW2X hosted in Burnaby. This machine was shared by other research groups and was accessed remotely, so timing data is dependent on unknown levels of network lag and job-queuing/load. All of the results reported in the paper are of the DW2X runs.

\subsection{Hyperparameter Results}
\begin{figure}
	\centering
	\includegraphics[width=0.7\linewidth]{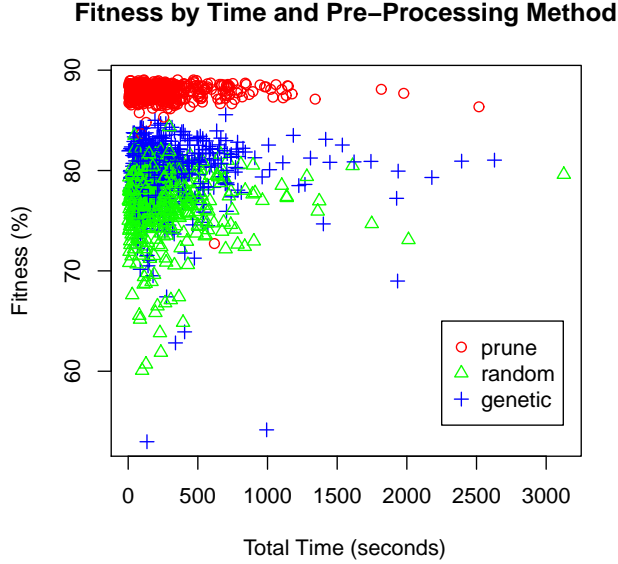}
	\caption{Plot of the coverage percentage vs. total processing time. Total time includes both actual pre and post-processing time, as well as network latency and D-Wave wait time. Color figures available in online version.}
	\label{fig:fitnesstimepreproccess}
\end{figure}

\begin{figure}
	\centering
	\includegraphics[width=0.7\linewidth]{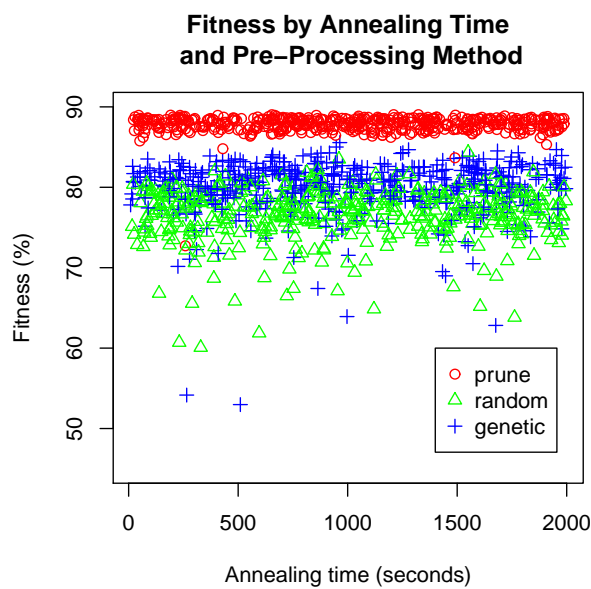}
	\caption{Plot of the coverage percentage vs. total time that the QA was responsible for. Total time neglects pre and post-processing time, but does include network latency, and D-Wave wait time, and the computational time needed to interpret the problem into QUBO form.  Color figures available in online version.}
	\label{fig:fitnessannealingandpreprocess}
\end{figure}

As discussed in the methods section, we randomly picked hyperparameter values to explore the full space. In general, the preprocessing stage was the most significant value, with clear differences in performance based on pruning, GA, and random choice, as can be seen in figures \ref{fig:fitnesstimepreproccess} and \ref{fig:fitnessannealingandpreprocess}. In these figures, total time and total annealing time increase with some combination of hyperparameters, such as NumGen or ProgTime, with fairly constant results, while there is a clear progression from Random to Genetic to Prune for the best results. As the pruning method involves outside knowledge of the problem domain, this is not incredibly surprising. Equally interesting, the GA run does appear to produce noticeable improvement.

\begin{figure}
\centering
\includegraphics[width=0.7\linewidth]{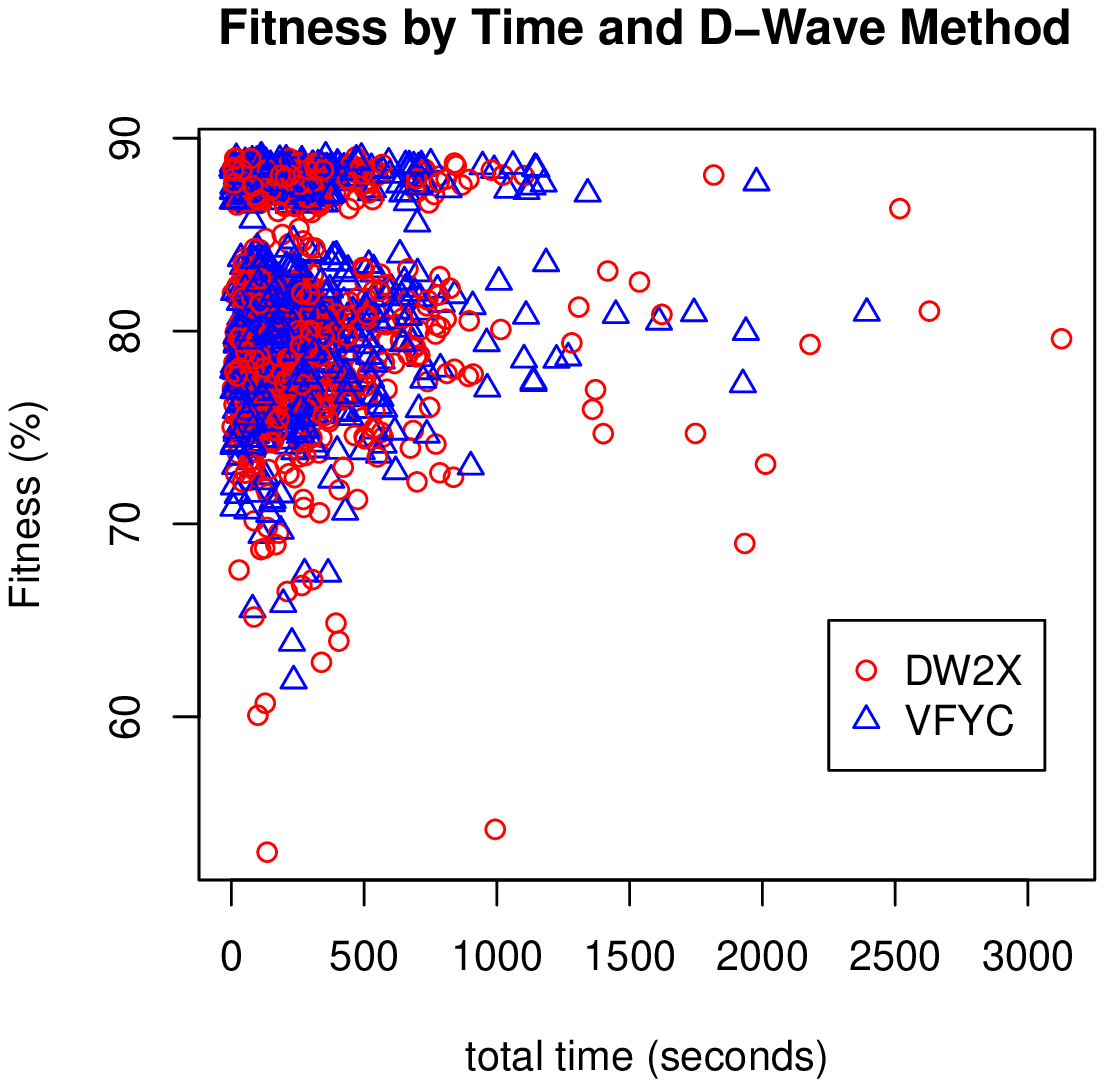}
\caption{Plot of the coverage percentage vs. total time, colored by which D-Wave processor was used. The DW2X is the actual processor, the Virtual Full Yield Chip (VFYC) has heuristics to simulate the results if all of the qubits and connectors had been fully functional. Color figures available in online version.}
\label{fig:fitnesstimeanddwavemethod}
\end{figure}

We also looked at the two D-Wave solvers, the real chip (DW2X), and the heuristically enhanced Virtual Full Yield (VFYC), as seen in Figure \ref{fig:fitnesstimeanddwavemethod}. The difference between the pre-processing methods dominates the difference, but regression analysis shows that there is a small but noticeable improvement when using the VFYC.


\subsection{Timing}
\begin{figure}
	\centering
	\includegraphics[width=0.7\linewidth]{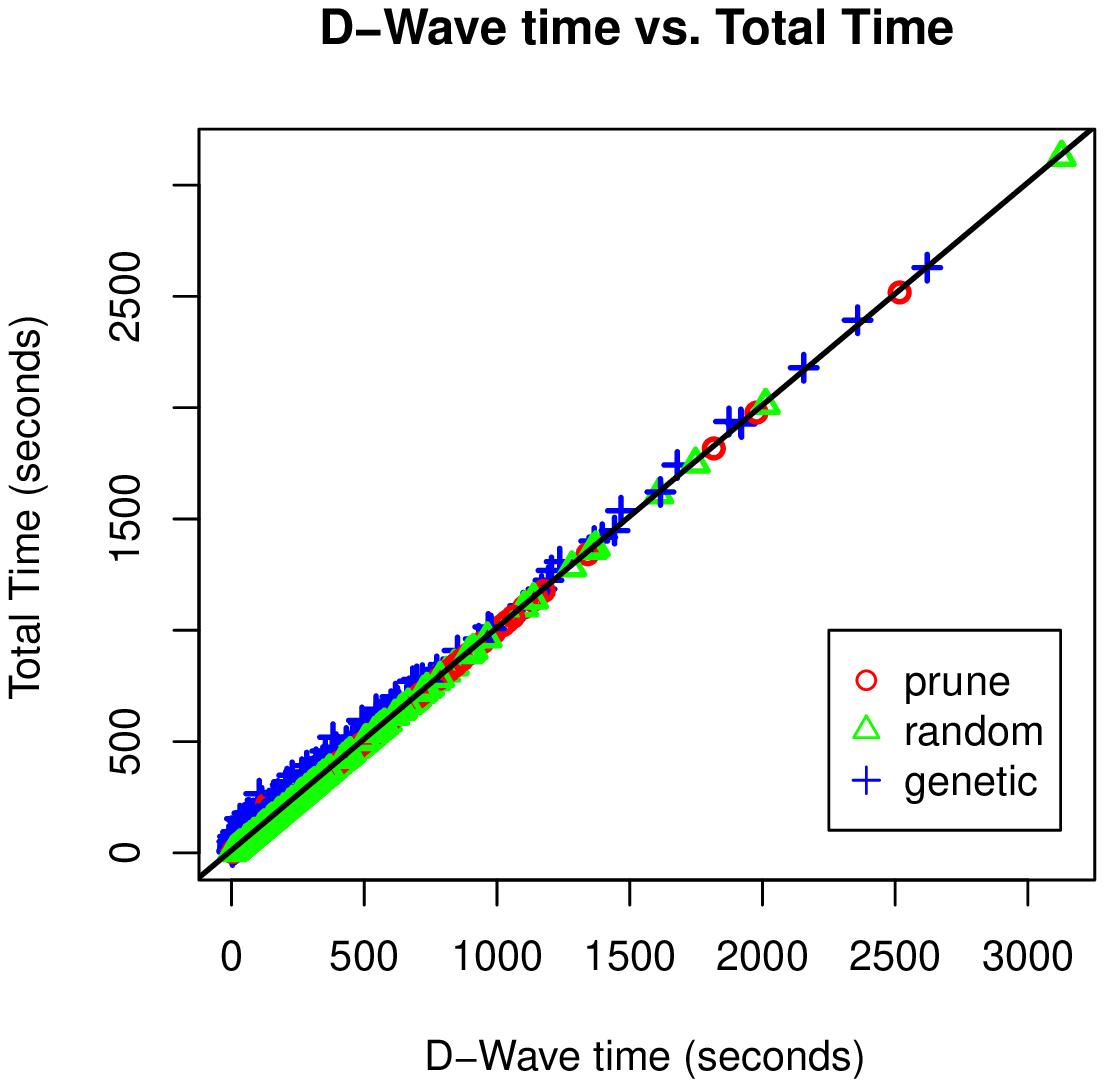}
	\caption[Total Time vs. D-Wave Time]{Most of the time spent was with some combination of D-Wave time. This time includes the actual annealing process, as well as network lag and job-queuing wait time. Color figures available in online version.}
	\label{fig:dwavetimevstotal}
\end{figure}

In optimization problems, the amount of time needed to produce results is of equal importance as the quality of results. Most heuristic algorithms can produce steady incremental improvements given arbitrarily large amounts of computational time and effort. In this work, both the pruning and random selection pre-processing methods were computationally trivial and took virtually no time. While the GA and post-processing common to all stages was a bit more computationally complex, the total analysis time was still dominated by the quantum annealing procedure (see figure \ref{fig:dwavetimevstotal}). 

A more rigorous comparison with other approaches would be misleading at best, as much of the time was not spent in actual wall-time and instead involved various other delays, particularly given inconsistent wait times for the D-Wave depending on utilization. 
	
\subsection{Regression Analysis}
As a final discussion on the importance of the various hyperparameters and pre-processing methods, we performed a simple linear least squares fitting regression. Each of the hyperparameters described in table \ref{tab:hyperparameters} were fed in as independent variables, with the exception of Method and Solver, which were changed to the binary variables: "Prune and Genetic" and "DW2X." Implicitly, when both Prune and Genetic are false, "Random" would be true. Similar reasoning also rules out the use of "VFYC", so both were not included in the regression results.

Table \ref{tab:Regression} shows the results of this regression. Both pre-processing methods were found to be significant at the 0.001 chance, as were the NumGen and NumNodes variables. As was clearly visible in the plots, Prune had the largest positive effect on coverage, with Genetic the next best. Increasing the number of generations had a positive effect, although increasing the population size had had a surprisingly negative effect. 

Increasing the number of nodes fed into the QA stage was also highly significant. This is important, as the limiting factor for this variable is the size of the quantum chip. As chips continue to scale rapidly, this indicates that future performance is likely to increase as well. 

Finally, we note that the DW2X chip had worse results than the VFYC. Through heuristics, the virtual full yield chip has all qubits fully working. This apparently resulted in an improvement, justifying this heuristic approach.

Also interesting is that the other annealing hyperparameters, including AnnealTime, ProgTime, and ReadTime, and even NumReps, had very small not statistically significant results. Essentially this means that good results are found even with very small values of all of these terms.

\begin{figure}
\centering
\includegraphics[width=0.7\linewidth]{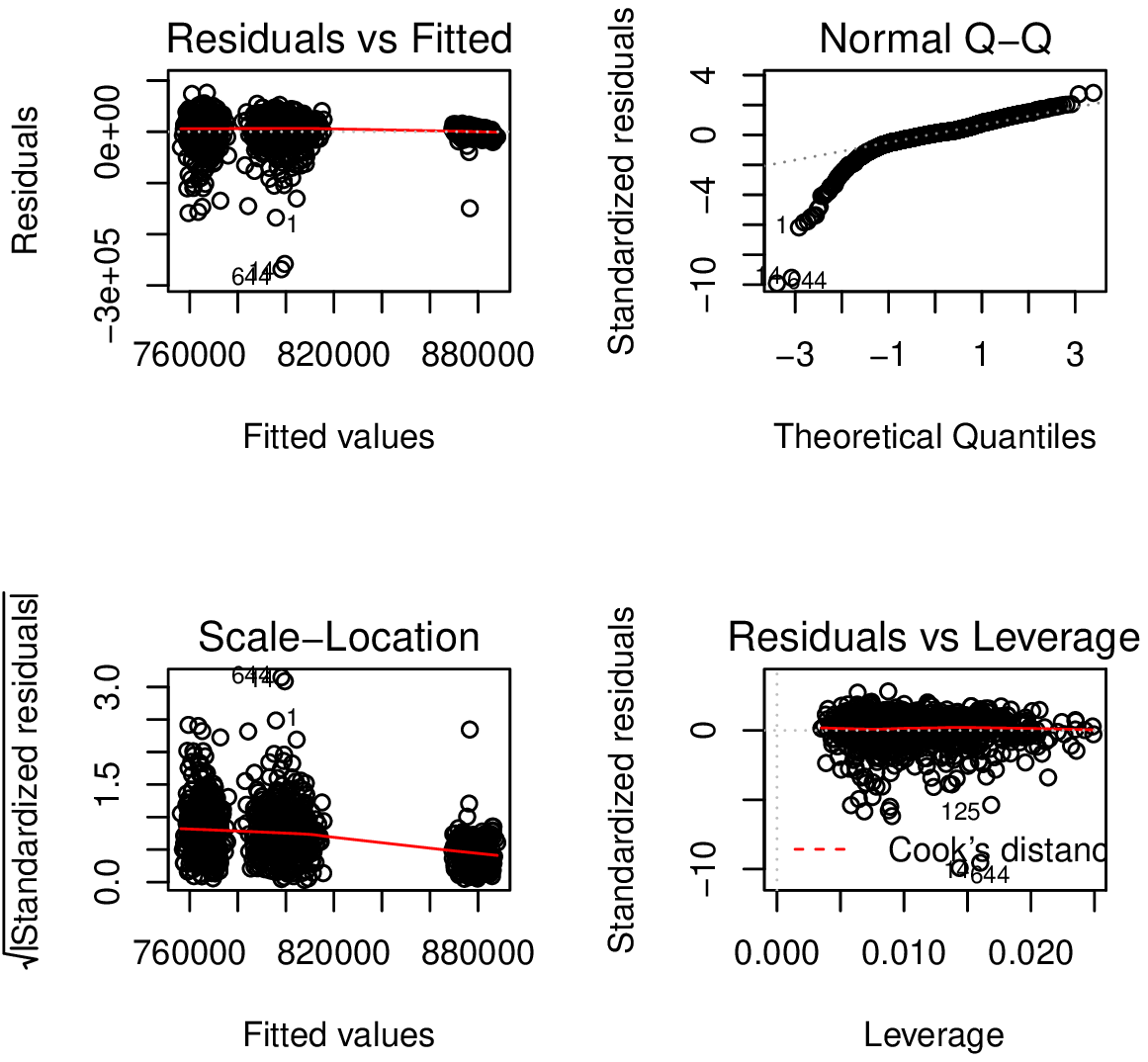}
\caption[RegressionAnalysis]{Plots for the regression analysis. Top Left: The three pre-processing methods each show their own clouds. In each group, error appears to be fairly evenly distributed around zero, with a long tail in the negative. Top Right: Quantile-Quantile plot. This mostly follows the 45 degree angle with a kink in the bottom right indicating some bi-modality. Bottom Left: Scale-Location plot. The horizontal fit indicates the data is homeoscedastic. Bottom Right: Residuals vs leverage plot. All points are well within Cook's Distance (which is outside the range of the plot), indicating there are no major outliers skewing the results of the regression.}
\label{fig:regression}
\end{figure}

Figure \ref{fig:regression} shows an analysis of these statistical results.

\begin{table}
	\caption{Regression Results}	
		\begin{center}
		Residual standard error: 27260 on 1413 degrees of freedom\newline 
		Multiple R-squared:  0.7519,	Adjusted R-squared:  0.7496\newline 
		F-statistic: 329.3 on 13 and 1413 DF,  p-value: $<$ 2.2e-16
		\end{center} 
		\label{tab:Regression}
	\begin{center}
		\begin{tabular}{|c|c|c|c|c|c|}
				\hline
                VarName &Estimate & Std. Error & t value &Pr($>|t|$) & Sig. \\    
                \hline
                (Intercept)    &7.457e+05 & 6.801e+03 &109.647 & $<$ 2e-16 &0.001\\
                Prune          &1.122e+05 & 1.791e+03 & 62.651 & $<$ 2e-16 &0.001\\
                Genetic        &2.793e+04 & 4.557e+03 &  6.129 &1.15e-09 &0.001\\
                Random$^\dagger$         &       NA &        NA &     NA &      NA &   \\
                NumGen       & 1.635e+01 & 4.169e+00 &  3.922 &9.20e-05 &0.001\\
                PopSize      &-8.177e+00 & 4.634e+00 & -1.764 & 0.07787 &0.1  \\                MutRate      & 9.303e+03 & 1.727e+04 &  0.539 & 0.59023 &   \\
                NumNodes      &5.013e+02 & 1.260e+02 &  3.980 &7.24e-05 &0.001\\
                LargestGroup &-5.364e+02 & 6.469e+02 & -0.829 & 0.40714 &   \\
                NumReps      & 2.189e-01 & 3.355e-01 &  0.652 & 0.51426 &   \\
                AnnealTime & 1.864e+00 & 1.273e+00 &  1.464 & 0.14342 &   \\
                ProgTime & 2.533e-01 & 2.431e-01 &  1.042 & 0.29758 &   \\
                ReadTime & 2.505e-01 & 2.445e-01 &  1.025 & 0.30568 &   \\
                SpinReverse   &-1.218e-02 & 4.406e-01 & -0.028 & 0.97795 &   \\
                DW2X          &-3.874e+03 & 1.453e+03 & -2.667 & 0.00775 &0.01 \\
                VFYC$^\dagger$          &        NA &        NA &     NA &      NA &   \\
                \hline
\end{tabular} 
\end{center}
$^\dagger$not defined due to singularity with other methods/solvers
\end{table}

\section{Conclusion}
Hard combinatorial optimization problems are intractable even for powerful modern computing systems, and will likely to remain so even given Moore's Law growth in computational power. Quantum computers, and QA in particular, offer a new paradigm to help overcome these problems. However, current generation QA hardware has many restrictions, and thus the most promising results come from a heterogeneous quantum/classical approach. We have developed preliminary results along these lines, working on a real-world problem. This research is valuable not only for improving solutions to this particular problem, but also as a tool for solving other hard optimization problems in many different contexts.

\section{acknowledgments}
We acknowledge the support of the Universities Space Research Association, Quantum
AI Lab Research Opportunity Program, Cycle 2, in particular for providing us with D-Wave access. We also thank the anonymous referees for many helpful suggestions for improving the article. We also acknowledge Brad Lackey for many helpful conversations regarding how to transform the problem into the QUBO form.

\bibliographystyle{unsrt}       

\bibliography{References}

\end{document}